# Towards More Efficient Shared Autonomous Mobility: A Learning-Based Fleet Repositioning Approach

Monika Filipovska, *Member, IEEE,* Michael F. Hyland, and Haimanti Bala

**Abstract** Shared-use autonomous mobility services (SAMS) present new opportunities for improving accessible and demand-responsive personal mobility. A fundamental challenge that SAMS face is appropriate positioning of idle fleet vehicles to meet future demand – a problem that strongly impacts service quality and efficiency. This paper formulates SAMS fleet repositioning as a Markov Decision Process and presents a reinforcement learning-based repositioning (RLR) approach called integrated system-agent repositioning (ISR). The ISR learns a scalable fleet repositioning strategy in an integrated manner: learning to respond to evolving demand patterns without explicit demand forecasting and to cooperate with optimization-based passenger-to-vehicle assignment. Numerical experiments are conducted using New York City taxi data and an agent-based simulation tool. The ISR is compared to an alternative RLR approach named externally guided repositioning (EGR) and a benchmark joint optimization (JO) for passenger-to-vehicle assignment and repositioning. The results demonstrate the RLR approaches' substantial reductions in passenger wait times, over 50%, relative to the JO approach. The ISR's ability to bypass demand forecasting is also demonstrated as it maintains comparable performance to EGR in terms of average metrics. The results also demonstrate the model's transferability to evolving conditions, including unseen demand patterns, extended operational periods, and changes in the assignment strategy.

## I. INTRODUCTION

THE rapid development of mobility-on-demand (MOD) services over the past decade has redefined personal mobility and is envisioned to create significant economic and social value by promoting demand-responsive and accessible multimodal transportation [1]. The inclusion of autonomous vehicles (AVs) in mobility service fleets open further opportunities for competitive performance and service quality of shared-use autonomous mobility services (SAMSs) and eliminate any costs and constraints associated with human drivers [2]. A fundamental challenge for efficient SAMS operation is the appropriate positioning of idle fleet vehicles to meet future demand, which has been shown to strongly influence both the quality of service and efficiency of MOD systems [3].

This paper focuses on the development of a learning-based fleet repositioning approach to significantly enhance the operational performance of a SAMS fleet in terms of service quality and efficiency. SAMS fleet repositioning is a complex problem that is challenging to formulate and solve due to several reasons that are addressed in this paper. The initial complexity of the task is due to the need to make coordinated and cooperative repositioning decisions for the entire fleet to avoid creating competition within the fleet. However, the outcomes of repositioning are further dependent on various system interactions and uncertainties. In short, the nature of the task requires that the repositioning decisions are appropriately suited to the corresponding vehicle-to-passenger matching process. Moreover, the repositioning strategy's performance will be impacted not only by the vehicle's final repositioning location, but also when and where they are available for passenger matching as they reposition. Finally, repositioning decisions need not only be predictive with respect to future demand for rides, but also anticipatory of future system states that occur due to interactions with other fleet decisions.

This work was supported in part by the Office of the Vice President for Research (OVPR) at the University of Connecticut.

M. Filipovska is with the Department of Civil and Environmental Engineering, University of Connecticut, Storrs, CT 06269 USA (e-mail: monika.filipovska@uconn.edu).

M. F. Hyland is with the Department of Civil and Environmental Engineering, University of California, Irvine, CA 92697 USA (e-mail: hylandm@uci.edu).

H. Bala is with the Department of Civil and Environmental Engineering, University of University of Connecticut, Storrs, CT 06269 USA (e-mail: haimanti.bala@uconn.edu)

To capture the above-mentioned considerations and complexities of the problem, this paper presents a reinforcement learning (RL) based approach for SAMS fleet repositioning. Envisioning a central fleet operator in a SAMS context, we formulate the fleet repositioning problem as a Markov Decision Process (MDP) with a single decision maker (operator). The fleet repositioning is modeled to interact with (i) an unknown demand-generating process and (ii) the operator's independent vehicle-to-passenger assignment decision. Therefore, the presented RL-based approach finds a repositioning solution that is anticipatory of how the system will evolve due to both (i) and (ii). The formulation and solution approach make use of the special characteristics of the SAMS fleet rebalancing problem that are often overlooked in the existing literature, as further described in the following sections.

### A. Literature Review

The literature on SAMS addresses problems in a variety of contexts and problem settings, and a comprehensive review by Narayanan et al. [4] covers various aspects beyond the scope of this study. Shared autonomous vehicles (SAV) services and autonomous mobility-on-demand (AMoD) are both terms that the literature often uses interchangeably with SAMS. As illustrated in Figure 1, the literature relevant to this study consists of two primary problem classes: (a) demand forecasting [5] and (b) fleet operations, further classified into (i) vehicle-passenger assignment (i.e., matching) [2], [6], [7] and (ii) idle vehicle management, typically repositioning [3], [8], [9].

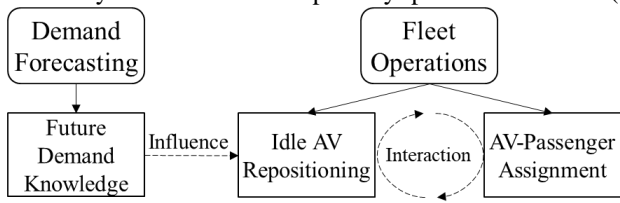

Fig. 1. Problem classes in SAMS literature

#### 1) SAMS Fleet Operational Problems

Assignment (i.e., matching) problems, determining the match between vehicles and passengers, are the most commonly studied in literature. They have been addressed in the context of taxi dispatching [7], for on-demand SAMS [6], and in more comprehensive contexts finding equilibria between service costs, waiting times, and demand [10]. Extensions that include shared rides have also emerged [11] and shown to have operational benefits [12]. Repositioning problems were introduced to manage idle vehicles within the service area in anticipation of future travel demand. These two sub-classes are highly interrelated since the purpose of repositioning is to position the vehicles at the right locations and times to facilitate better assignment. A prominent issue with fleet repositioning is the increase in vehicle miles travelled by empty vehicles [9], [13]. An early study [14] considers vehicle rebalancing as a linear programming problem and develops a real-time rebalancing policy suited for highly variable environments. Later, a queueing-based formulation of the problem was presented with a non-myopic solution using a Lagrangian decomposition heuristic [3]. The basis for the repositioning decision in these studies is knowledge of the random process that generates the demand.

#### 2) Demand Forecasting for SAMS

Demand forecasting for SAMS is a less prominent class of problems, categorized into offline methods using historical data and online methods relying on and real-time data access [5]. Short-term demand forecasting has been studied in the context of carsharing [15], taxis [16], [17], and public transportation services [18]. Despite being studied as two separate problem classes, the SAMS operational and the forecasting problems are inherently interconnected, as demonstrated by [19], among other studies.

#### 3) RL for SAMS Fleet Operations

Reinforcement learning (RL) methods have recently emerged as an alternative for solving SAMS operational problems, since they can be employed to devise efficient algorithms for complex decision making, often without significantly compromising optimality. A recent survey paper reviews the use of RL for several problem classes including pricing, matching, vehicle repositioning, routing, and shared rides (i.e., ride-pooling) [20]. Special-context problems have also explored the use of RL, including ride-hailing with electric vehicles [21], shared rides [22], and meta-learning for transferability [23].

RL-based repositioning (RLR) approaches fall into one of two categories, according to [20], based on whether the RL agent models (a) the operator (i.e., service provider) or (b) the vehicles. The latter is more common in the literature, where the vehicle agents are coordinated using multi-agent reinforcement learning (MARL) modeling a joint action space for all vehicles in the fleet [24], [25], [26], [27], [28]. The drawback is that the problem becomes intractable due to the quickly increasing action space with larger fleet sizes. To overcome this challenge, a few studies have proposed decentralized approaches for vehicle agents. For example, [29] presents an approach where each vehicle autonomously learns its behavior for repositioning and assignment. The challenge with such decentralized approaches, however, is that non-cooperative repositioning decisions lead to competing actions by multiple vehicles and result in a supply surplus in high-demand areas [20].

To overcome these challenges, single-agent RL formulations have recently emerged [20], [30] where the system or operator is modeled as an agent. For example, [30], [31] present scalable approaches for modeling the SAMS operator as a single agent making a multi-dimensional decision for the entire fleet. In these formulations, however, the RL agent directly observes future demand as part of its system state, which makes it challenging to evaluate the performance of the resulting strategy. Furthermore, the existing formulations limit the interactions of repositioning decisions with passenger-vehicle assignment in several ways due to how they model the repositioning decisions. Nevertheless, these studies demonstrate the potential of single-agent RLR approaches in this context.

#### 4) Research Gaps

The recent literature demonstrates the potential of RL approaches for SAMS fleet repositioning problems. The findings show that vehicle-agent approaches have scalability issues if using MARL and cooperation otherwise. Recent developments reveal that system-agent RL formulations resolve both issues by enabling scalable solution approaches for coordinated decisions in large SAMS fleets. However, system-agent RL formulations are rare in the literature, and the existing approaches are relatively rudimentary compared to corresponding vehicle-agent formulations. The problem formulations do not account for the complex interactions of repositioning decisions with other operator decisions, such as passenger-vehicle assignment. Furthermore, their dependence on knowledge of future demand makes them unsuitable for practical applications and difficult to evaluate. To better understand the potential of system-agent RL approaches there is a need to evaluate their performance without explicit knowledge of future demand and to assess their potential if better integrated with the assignment decision. The limitations of prior system-agent studies also reveal the need for more testing and understanding of the transferability of RLR solutions in an evolving transportation system.

### B. Paper Contributions

Given these research gaps, this paper presents a novel, integrated system-agent RL-based repositioning (ISR) approach for SAMS fleet repositioning with several contributions to the literature:

1. The system-agent repositioning decision is formulated to allow for interaction with the operator's assignment decisions, not accounted for in existing studies.

2. The repositioning decision is learned with no input of the system's demand-generating process.

3. The RL approach learns a repositioning decision that jointly interacts with the request arrival and the traveler-to-vehicle assignment process, resulting in an integrated solution framework for centralized SAMS decisions.

4. The system agent's multi-dimensional action is defined to allow for sufficient specificity while ensuring that the problem size does not increase with the size of the vehicle fleet, but only with the spatial disaggregation of the service area. This is a significant advantage relative to MARL approaches whose performance degrades substantially with increasing vehicle fleet sizes.

5. Numerical experiments demonstrate the performance of the presented approach and its transferability to unseen demand patterns and new assignment strategies that are not encountered in the model's training.

The remainder of this paper is organized as follows. Section II presents the problem setting and definition, followed by the solution framework in Section III. Numerical experiments are presented in Section IV, and the paper is concluded in Section V.

## II. PROBLEM SETTING AND DEFINITION

### A. Problem Setting with Notation

This paper focuses on a problem where a SAMS central operator controls a fleet of autonomous vehicles (AVs) in the set $V = \{1, 2, \ldots, |V|\}$ to serve trip requests from users $R = \{1, 2, \ldots, |R|\}$ that arrive dynamically over time. The problem is restricted to a finite time horizon $T = [0, t^*]$ and a pre-defined service area to which all events and decisions are constrained.

These system components with their notation, definitions, are characteristics are summarized in Table I. The service area is separated into non-overlapping sub-areas for the purpose of defining the repositioning problem, but locations of passengers and vehicles are not discretized and can be at any point within the service region. The key information on passengers and vehicles is the definition of states $s_r(t) \forall r \in R, t \in T$ and $q_v(t) \forall v \in V, t \in T$ for passengers and vehicles, respectively. The state definitions are used to define mutually exclusive and collectively exhaustive subsets of requests and vehicles based on their states, which will be used in the problem formulations.

The request state constraints ensure that the state for any request is fixed to "unrequested" while $t < t_r$ and becomes "unassigned" $t = t_r \forall r \in R$. Under the assumption that travelers cannot cancel their requests and will continue to wait until assigned, the monotone increasing property of $s_r(t)$ over time ensures that a request has to change states in order from unrequested to assigned to in-vehicle and then become served. While the problem definition can accommodate a constraint on maximum wait time after which travelers cancel their requests, here we focus on evaluating and observing the full range of potential waiting times in a more general sense as an indicator of service quality. Finite wait time constraints can be meaningful for evaluating the cost of providing service relative to the price of a ride given demand elasticity.

The vehicle state definitions assume that vehicles are either idle or performing one of three possible tasks: en-route passenger pick-up, en-route drop-off, or repositioning. In this problem we assume AVs do not need to refuel within the analysis period. The constraints on vehicle state transitions assume that vehicles will be serving one passenger at a time and cannot drop a passenger after being assigned to it until the passenger has been picked up and dropped off at the destination.

### B. Problem Definition for SAMS Operator Decisions

In this paper, we are investigating a SAMS system in which a single operator controls a fleet of vehicles by making two types of operational decisions: passenger-vehicle assignment (primary decisions) and vehicle repositioning (secondary decisions). We assume the operator makes decisions with the goal of operating the service efficiently and providing high service quality for passengers. In these operational decisions we assume a simple ride-hailing system, where ride sharing, dropping, switching, or hopping are not allowed. The decisions are constrained so that the

| | Notation | Definition | Characteristics |
|---|---|---|---|
| | $T$ | Set of time intervals | $t \in T$ denotes a time interval $t$ in the set |
| | $K$ | Set of sub-areas of the service region | $k \in K$ a sub-area in the set; non-overlapping: $k_1 \cap k_2 = \emptyset \ \forall \ k_1 \neq k_2 \in K$ |
| **Passenger Requests** | $R$ | Set of passenger requests | $r \in R$ a request $r$ in the set |
| | $t_r$ | Request time for $r \in R$ | $t_r \in T \ \forall \ r \in R$ |
| | $o_r, d_r$ | Origin and destination locations for $r \in R$ | $o_r, d_r \in \bigcup_{k \in A} k \ \forall \ r \in R$ |
| | $p_r(t)$ | Position (physical location) of request $r \in R$ at time $t \in T$ | $p_r(t) \in \bigcup_{k \in A} k \ \forall \ r \in R, t \in T$; $p_r(t) = o_r \forall \ r \in R, t \leq t_r$ i.e., position must be fixed to the request origin until the request time. |
| | $s_r(t)$ | State of a request $r$ at time $t \ \forall \ r \in R, t \in T$ | $s_r(t) \in \{0,1,2,3,4\}$ denoting unrequested, unassigned, assigned, in-vehicle, and served requests, respectively. $s_r(t)$ is monotone increasing over $t$: i.e., for $t_1, t_2 \in T$ and $t_2 > t_1$, $s_r(t_2) \geq s_r(t_1) \forall \ r \in R$ $s_r(t) = 0$ if $t < t_r \ \forall r \in R$ and $s_r(t_r) = 1 \ \forall \ r \in R$ If $s_r(t) = 4$ then $p_r(t) = d_r \ \forall \ r \in R$, i.e., position must be fixed to the request destination after drop-off. |
| | $R_t \subseteq R$ | Known requests at $t \in T$ | $R_t = \{r \in R \ | t_r \leq t\} = \{r \in R | s_r(t) \neq 0\}$ |
| | $R_U \subseteq R$ $R_A \subseteq R$ $R_{IV} \subseteq R$ $R_S \subseteq R$ | Subsets of requests by state: $U$ = unassigned, $A$ = assigned, $IV$ = in vehicle, $S$ = served | $R_U = \{r \in R_t | s_r(t) = 1\}$; $R_A = \{r \in R_t | s_r(t) = 2\}$ $R_{IV} = \{r \in R_t | s_r(t) = 3\}$; $R_S = \{r \in R_t | s_r(t) = 4\}$ Mutually exclusive and collectively exhaustive: $R_i(t) \cap R_j(t) = \emptyset \ \forall \ i \neq j \in \{U, A, IV, S\} \ \forall \ t \in T$ $R_U(t) \cup R_A(t) \cup R_{IV}(t) \cup R_S(t) = R_t \ \forall \ t \in T$ |
| | $w_r(t)$ | Elapsed wait time for request $r \in R_t$ at $t \in T$ | $w_r(t) = \begin{cases} t - t_r & \text{if } s_r(t) \leq 2 \\ t'_r - t_r & \text{otherwise} \end{cases}$, where $t'_r$ denotes the pick-up time for request $r$ if $s_r(t) > 2$: $t'_r = \max(t \in T \ | \ s_r(t) \leq 2)$ |
| **Vehicle Fleet** | $V$ | Set of AVs | $v \in V$ denotes a vehicle (AV) in the set |
| | $l_v(t)$ | Location $\forall v \in V, t \in T$ | $l_v(t) \in \bigcup_{k \in K} k \ \forall \ v \in V, t \in T$ |
| | $q_v(t)$ | State of vehicle $v \in V$ at $t \in T$ | $q_v(t) \in \{1,2,3,4\}$ denoting idle, en-route pick-up, en-route drop-off, and repositioning AV, respectively. Transitions between states for consecutive $t_1 < t_2$: $\quad q_v(t_1) \in \{1,4\} \implies q_v(t_2) \in \{1,2,4\}$ $\quad q_v(t_1) = 2 \implies q_v(t_2) \in \{2,3\}$ $\quad q_v(t_1) = 3 \implies q_v(t_2) \in \{3,1\}$ |
| | $V_I \subseteq V$ $V_P \subseteq V$ $V_D \subseteq V$ $V_R \subseteq V$ | Subsets of vehicles by state: $I$ = idle, $P$ = en-route pick-up, $D$ = en-route drop-off, $R$ = repositioning | $V_I(t) = \{v \in V | q_v(t) = 1\}$; $V_P(t) = \{v \in V | q_v(t) = 2\}$ $V_D(t) = \{v \in V | q_v(t) = 3\}$; $V_R(t) = \{v \in V | q_v(t) = 4\}$ Mutually exclusive and collectively exhaustive: $V_i(t) \cap V_j(t) = \emptyset \ \forall \ i \neq j \in \{I, P, D, R\} \ \forall \ t \in T$ $V_I(t) \cup V_P(t) \cup V_D(t) \cup V_R(t) = V \ \forall \ t \in T$ |

operator does not reject any traveler requests but only optimizes when and how to serve them. Furthermore, in focusing on the operational side, we do not account for strategic or tactical decisions, such as fleet size optimization, service pricing, or cost-revenue evaluation.

In primary decision making, referred to as the matching or assignment problem, the operator aims to reduce operational costs and increase service quality, measured via fleet miles travelled and traveler wait times, respectively. In secondary decision making, the vehicle repositioning problem entails choosing where to direct idle vehicles to enable better primary decisions. Hence, idle vehicle repositioning occurs to support the primary decisions by anticipating where and when vehicles will be needed for assignment. Due to this, AVs in repositioning will be considered available for assignment in the primary decision stage. Focusing on the idle AV repositioning (secondary decision), this paper aims to solve this problem for a system where changes in the system occur due to incoming ride requests (stochastic demand) and due to the operator's passenger-vehicle assignment decisions. Thus, we first formulate the AV repositioning problem, which is central to this paper, and then turn to the assignment problem.

### 1) AV Repositioning Problem Formulation

The AV repositioning problem is defined as a fully observed Markov Decision Process (MDP) $\mathcal{M}_{reb}(\mathcal{S}^{reb}, \mathcal{A}^{reb}, p^{reb}, r^{reb})$ consisting of the state space, action space, probabilistic dynamics, and reward function, respectively. The aim is to learn a behavioral policy that will make the best decision for repositioning of vehicles between sub-areas. The service area is represented as a graph $\mathcal{G}(\mathcal{V}, \mathcal{E})$ where the vertices $\mathcal{V}$ are the sub-areas $k \in K$ as defined above, represented by their centroids, and the edges $\mathcal{E}$ represent shortest paths connecting the sub-areas' centroids as an approximation of travel times between them. $\mathcal{G}$ is a complete directed graph, where $\mathcal{E} = \{(i,j) \forall i, j \in \mathcal{V}\}$. In the discrete time horizon $T$, the state $S^t \in \mathcal{S}^{reb} \ \forall \ t \in T$ contains the information needed to determine the repositioning strategy, the action $A^t \in \mathcal{A}^{reb} \forall t \in T$ is a behavior policy that describes a distribution of the vehicles over the edges of $\mathcal{G}$. $P^{reb}$ describes the dynamics of the system as it evolves to the next state through a conditional probability distribution: $P^{reb}(S^{t+1}|S^t, A^t)$ given the state and action in the previous time step. Finally, $r^{reb}: S \times A \to \mathbb{R}$ is a reward function that should guide the system to desirable states. Each of these components are described in more detail below.

In modeling the vehicle repositioning problem, we assume that only idle AVs can be given the task of repositioning, and they will reposition from their current sub-are to another sub-area. Our formulation does not accommodate repositioning to a new location within the same sub-area, but we allow a vehicle to remain idle and not be repositioned altogether.

The state $S^t$ contains knowledge of the graph $\mathcal{G}(\mathcal{V}, \mathcal{E})$ along with vertex- and edge-level information by means of a feature matrix $\mathbf{N}$ and adjacency matrix $\mathbf{M}$. $\mathbf{N}(t) \forall t \in T$ contains the following information for each sub-area represented by a vertex: the numbers of idle vehicles ($c_i^{idle}(t) = |\{v \in V_I(t)|l_v(t) \in i\}| \forall i \in \mathcal{V})$, incoming repositioning vehicles: ($c_i^{rep}(t) = |\{v \in V_R(t)|d_v(t) \in i\}| \forall i \in \mathcal{V})$, arriving passenger-carrying vehicles ($c_i^{arr}(t) = |[v \in V_D(t)|d_r \cdot y_{rv}(t) \in k]| \forall i \in \mathcal{V})$, originating passenger requests for a specified number of $q$ past time periods ($c_i^{pass}(t') = |[r \in R(t)|o_r \in i, r_t \in [t'-1, t']| \forall i \in \mathcal{V}, t' \in \{t - q + 1, \dots, t - 1, t\}])$. The state only includes observations of total demand in each sub-area (vertex) for $q$ past time intervals, but not for future time intervals as commonly seen in the literature. The adjacency matrix $\mathbf{M}$ describes the graph $\mathcal{G}(\mathcal{V}, \mathcal{E})$ via the edge lengths, i.e., pairwise travel times between the sub-areas' centroids. We assume fixed shortest-path travel times on a Manhattan grid, but the formulation allows for a time-varying $\mathbf{M}(t) \forall t \in T$.

The action $A^t \in \mathcal{A}^{reb}$ is defined as a multi-dimensional action over the edges of $\mathcal{G}$, such that $A^t = [A_{ij}^t] \forall (i,j) \in \mathcal{E}$. The value for each dimension $A_{ij}^t \in [0,1]$ is the fraction of currently idle vehicles at vertex (i.e., sub-area) $i \in \mathcal{V}$ to be rebalanced toward vertex (i.e., sub-area) $j \in \mathcal{V}$ with the constraint that $\sum_{j \in \mathcal{V}} A_{ij}^t = 1 \forall i \in \mathcal{V} t \in T$. In this way, the action is not constrained by the size of the vehicle fleet and the magnitude of the action representing a fraction does not have to vary with the size of the fleet or the number of idle vehicles over time. In our implementation, the action is easily converted to the number of vehicles to be repositioned as $\rho_{ij}^t = [A_{ij}^t c_i^{idle}(t)]$ where $\lfloor \cdot \rfloor$ is the floor function ensuring $\rho_{ij}^t$ is an integer and less than or equal to the available number of vehicles $c_i^{idle}(t)$, at vertex (sub-area) $i$ and time $t$. Therefore, the goal in solving the MDP is to find an appropriate policy that defines a distribution over possible actions given states $\pi(A^t|S^t)$. $P^{reb}$ describes the system dynamics by capturing the evolution of the state over time. In this set up, that includes the evolution of the passenger demand patterns and the influence of primary and secondary decisions on the future state elements. Due to the stochastic demand, independent of the operator's decisions, $P^{reb}$ is an unknown stochastic process in this MDP.

The reward function is defined from the perspective of the operator to capture the system efficiency and service quality, aiming to approximate the mean passenger waiting time. However, since the mean is not a cumulative summative value, we express the reward at each time step via a negative (penalty) term for passenger waiting times and a positive term for the number of served passengers, with weights on each term. The reward at time $t$, with a time-discretization interval $\delta$, is computed as:

$$r^{reb}(t) = -\omega \cdot \delta \cdot (|R_U(t)| + |R_S^A(t)|) + \sigma |R_S^A(t)|, \quad (1)$$

where $R_U(t)$ is the set of unassigned requests (i.e., waiting passengers) at time $t$ and $R_S^A(t)$ is the set of requests served in the last time interval between $t - 1$ and $t$. Thus, $(|R_U(t)| + |R_S^A(t)|)$ passengers have been waiting during the last time interval of duration $\delta$ with a total wait time of $\delta \cdot (|R_U(t)| + |R_S^A(t)|)$. The second term is number of served requests during the past time interval $|R_S^A(t)|$, which can also be computed $|R_S^A(t)| = |R_S(t)| - |R_S(t-1)|$ since the parameters $\omega$ and $\sigma$ are weights that will be adjusted so that this reward approximates total mean wait time and are discussed below in more detail.

In determining the optimal policy, the goal is to optimize the decisions over the entire time horizon since the effects of repositioning decisions on the system state and performance are not immediate. Therefore, the objective is based on maximizing a cumulative reward resulting from all states and actions. For this purpose, we use the concept of the system's trajectory over a time horizon $T$, i.e., the sequence of states and actions: $\mathcal{T} = (S^{t=0}, A^{t=0}, S^{t=1}, A^{t=1}, \dots, S^{t=t^*}, A^{t=t^*})$. Given the stochastic system dynamics $P^{reb}$, the trajectory $\mathcal{T}$ under a policy $\pi$ has the distribution:

$$p_{\pi(\mathcal{T})} = d_0(s_0) \prod_{t=0}^{t^*} [\pi(A^t|S^t) P^{reb}(S^{t+1}|S^t, A^t)]. \quad (2)$$

Finally, the RL objective $J(\pi)$, given in (3), is to maximize the expected cumulative reward under $p_{\pi(\mathcal{T})}$:

$$\max_\pi J(\pi) = \max_\pi \mathbb{E}_{\mathcal{T} \sim p_{\pi(\mathcal{T})}} [\sum_{t=0}^{t^*} \gamma^t r^{reb}(S^t, A^t)], \quad (3)$$

where, $\gamma^t \in (0,1] \forall t \in T$ is a discount factor for the cumulative reward that can vary over time if needed, and $r(S^t, A^t)$ is the reward function. In our formulation, $\gamma^t$ is constant but could be modified to emphasize the importance of rewards obtained during important decision periods, such as peak demand periods. The reward weights were determined so that $J(\pi)$ approximates the negative mean request wait time and can be derived via the following steps. First notice that the cumulative reward can be expressed as:

$$\sum_{t=0}^{t^*} r = -\omega(\sum_{t=0}^{t^*} \delta \cdot (|R_U(t)| + |R_S^A(t)|)) + \sigma |R_S(t^*)|$$
$$= -\omega \cdot \sum_{r \in R_{t^*}} w_r(t^*) + \sigma |R_S(t^*)|. \quad (4)$$

Setting (4) equal to the desired reward value $-\sum_{r \in R_{t^*}} w_r(t^*)/|R_{t^*}|$ and $\omega + \sigma = 1$, results in:

$$\omega = \frac{\sum_{r \in R_{t^*}} w_r(t^*) - |R_S(t^*)||R_{t^*}|}{|R_{t^*}|(\sum_{r \in R_{t^*}} w_r(t^*) - |R_S(t^*)|)}, \quad (5)$$

which gives an exact solution for the weights. In practice the total number of requests, served requests, and total wait time must be approximated.

### 2) AV-Request Assignment Problem Formulation

The primary decision, AV-request assignment, is defined as a stochastic sequential optimization problem. This formulation is based on that presented by [6], since it is not the primary focus of this paper. The sequential matching problem is addressed by solving an AV-request assignment optimization problem at each decision interval over the analysis period. In each decision interval, assignment is made for the subset of unassigned requests $R' = R_U$ and available vehicles that are idle or relocating $V' = V_I \cup V_R$. Let $d_{rv}(t)$ denote the distance between the pickup location $o_r$ for traveler $r \in R$ and the location $l_v(t)$ of vehicle $v \in V$ at time $t$. The problem is formulated in two ways, according to [6]. First, if the number of requests is greater than the number of AVs available, i.e., $|R'| > |V'|$:

$$\min_{x_{rv}} \sum_{r \in R'} \sum_{v \in V'} (d_{rv} x_{rv} - \alpha w_r x_{rv}) \quad (6)$$

$$\text{s.t. } \sum_{v \in V'} x_{rv} \leq 1 \ \forall \ r \in R' \quad (7)$$

$$\sum_{r \in R'} x_{rv} = 1 \ \forall \ v \in V' \quad (8)$$

$$x_{rv} \geq 0 \ \forall \ r \in R', v \in V' \quad (9)$$

The objective function in (6) has two terms representing the total distance between travelers and AVs assigned and the elapsed wait time for assigned travelers. The weight $\alpha$ scales and converts the wait time units into the distance equivalent. The second formulation is for the case when $|R'| \leq |V'|$, where all requests can be assigned and they do not need to be prioritized according to waiting time, hence the objective function is: $\min_{x_{rv}} \sum_{r \in R'} \sum_{v \in V'} d_{rv} x_{rv}$. The constraints are parallel to the first formulation, with equality in equation (7) and inequality in equation (8) to model the excess of vehicles instead of excess of trip requests.

In the primary decision we assume that AVs maintain at most one request match at any given time. We model pickup and drop-off time as a short amount of time for passenger boarding and alighting, which does not impact the passengers' wait time. The specific values used for these parameters are specified in the numerical experiments. Vehicles are considered available for matching if they have no request match at a given time, i.e., they can be matched while idle or while performing a repositioning task.

## III. Solution Approach

The solution method consists of solving both operational problems defined above with a focus on the repositioning decisions. The operational decision-making process is implemented with an agent-based simulation tool for evaluation. The simulation tool models the operation of a SAMS fleet, including generating or obtaining demand for rides from external data, implementing the matching and repositioning decisions approaches, and simulating the movement of vehicles as they fulfill their tasks. The simulation tool used in this study has been presented and used in several studies in the literature [6], [19], [32], [33] and its functionality is summarized in Figure 2. The operational decision procedure chooses the next operational decision for the system, whether that be vehicle assignment or repositioning, or both decisions jointly or sequentially. The methodology for this step is the focus of this section.

The two-part SAMS operational framework, summarized in **Error! Reference source not found.**, solves the two decision problems, where the *SAMS System* contains all steps from the agent-based simulation except the operational decision procedure. The external demand generation is an input to the SAMS system, and the system delivers and receives information to and from the two decision-making components. Notably the two decision modules do not communicate with one another, but rather observe one another's impacts on the system state.

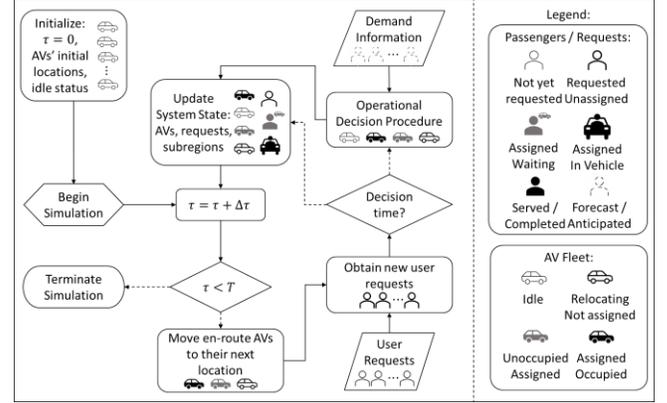

Fig. 1. Agent-based SAMS simulation framework

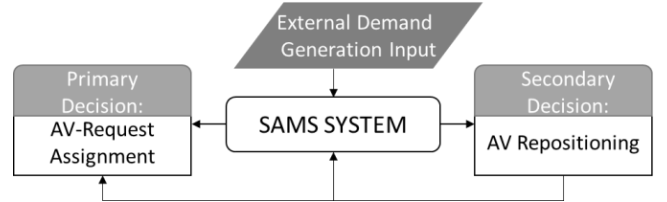

Fig. 3. Two-part operational framework for the SAMS system

### A. AV Repositioning Approach

Having defined the SAMS repositioning problem as an MDP, we solve it using an RLR approach that follows a standard learning loop: the system agent interacts with the environment $\mathcal{M}_{reb}$ by applying a behavior policy $\pi(A|S)$, where it observes the state $S^t$, chooses an action $A^t$, then observes the next state $S^{t+1}$ with a scalar reward feedback $r(t) = r(S^t, A^t)$. The procedure repeats and the agent uses the observed transitions $(S^t, A^t, S^{t+1}, r(t))$ to update the policy $\pi$. Therefore, we will refer to this approach as integrated system-agent RL-based repositioning (ISR).

The proposed solution approach estimates the gradient of the objective $J(\pi)$ and using an approximate gradient ascent approach to maximize it. We parametrize the policy $\pi$ by a parameter vector $\theta$ as $\pi_\theta(A^t|S^t)$, and the gradient of the objective $\nabla_\theta J(\pi_\theta)$ is expressed as:

$$\mathbb{E}_{\mathcal{T} \sim p_{\pi_\theta}(\tau)} \left[ \sum_{t=0}^{t^*} \gamma^t \nabla_\theta \log \pi_\theta(A^t|S^t) \hat{A}(S^t, A^t) \right], \quad (10)$$

where $\hat{A}$ is the advantage estimator:

$$\hat{A}(S^t, A^t) = \sum_{t'=t}^{t^*} \gamma^{(t'-t)} r(S^{t'}, A^{t'}) - b(S^t), \quad (11)$$

where $b(S^t)$ is an estimator for the state value function [34]:

$$V^\pi(S^t) = \mathbb{E}_{\mathcal{T} \sim p_{\pi_\theta}(\tau|S^t)} \sum_{t'=t}^{t^*} \gamma^{(t'-t)} r(S^{t'}, A^{t'}). \quad (12)$$

To solve this problem, we use the Advantage Actor-Critic (A2C) algorithm to determine the policy $\pi_\theta(A^t|S^t)$ and the value function estimator $V(S^t)$. A2C is an actor-critic (AC) method, which belongs to the temporal-difference (TD) class of methods that learn from raw experience, without a model of the environment dynamics. This is an important feature for the SAMS repositioning problem where the environment dynamics are unknown. AC methods separately parametrize the policy and the value function, so they are composed of a policy structure that selects the actions, known as the *actor*, and an estimated value function that evaluates (or criticizes) the actions, known as the *critic*. The *critic* learns and criticizes the policy that the *actor* follows, where the critique is parametrized via a TD error [35], [36]. A2C expands on this idea by employing a *critic* that estimates the advantage function $\hat{A}(S^t, A^t)$, instead of just the value function, to perform gradient ascent on the objective function $J(\pi_\theta)$ [34].

The A2C process for the SAMS repositioning problem is shown in Figure 4, where the "environment" contains all remaining components of the operational framework: the SAMS System, the primary decision, and the demand input. The A2C agent interacts with the environment by drawing observations of the environment state, observed by the *actor* and *critic* components. The *actor* develops a policy to choose an action feeding back into the SAMS system which implements changes resulting from this action. Based on the updated environment, a reward is observed by the *critic*, which then computes the advantage of the action and feeds it back to the *actor* to update the policy.

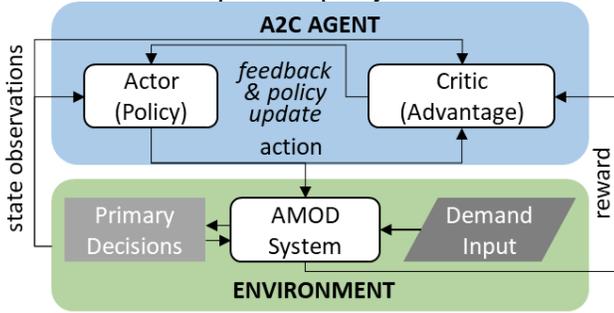

Fig. 4. A2C Agent with a SAMS system environment.

The A2C agent parametrizes the policy and the value-function estimator via deep neural network architectures for the *actor* and *critic* components, respectively. The policy defines the repositioning action $A^t = \left[A^t_{ij}\right] \forall (i,j) \in \mathcal{E}$, where $A^t_{ij} \in [0,1]$ specifies the percentage of currently idle vehicles at $i$ to reposition towards $j \in \mathcal{V}$, and $\sum_{j \in \mathcal{V}} A^t_{ij} = 1 \forall i \in \mathcal{V}$ $t \in T$. For $\pi_\theta(A^t|S^t)$ to define a valid probability density over actions, the output of the policy network represents the concentration parameters $\alpha$ of a Dirichlet distribution $Dir(\cdot)$, such that $A^t \sim Dir(A^t|\alpha) = \pi_\theta(A^t|S^t)$. The neural network for the actor consists of a graph attention network (GAT) layer, followed by four layers of graph convolutional neural network (GCN) with skip-connections and Rectified Linear Unit (ReLU) activations. Its output is aggregated across neighboring nodes using a permutation-invariant sum-pooling function and passed to three multi-layer perceptron (MLP) layers to produce the Dirichlet concentration parameters. The value function is defined using a similar architecture, where the main difference is a global sum-pooling performed on the output of the graph convolution so that it computes a single value function estimate for the entire network.

While the A2C algorithm is not new to the RL literature, the innovative aspect of the proposed implementation is that we tailor the A2C to exploit special features of the problem. Firstly, in defining the graph $\mathcal{G}(\mathcal{V}, \mathcal{E})$ based on the regions in the service area $k \in A$, the multi-dimensional action of A2C is defined over the edges of the graph. The graph nodes carry information about the system state including the number of passenger requests and vehicles in each region, while the edges carry information about region-to-region travel times. Secondly, the graph structure is the basis for the neural network architecture of the actor and critic, using GAT and GCN layers tailored for learning on graphs. The GAT layers use an attention mechanism to determine the strength of relations between regions as weights on the graph edges, while using the travel time and state information. Since deep networks of GAT layers are computationally expensive, we pair the GAT layer with GCN layers which perform deep learning on the graph structure output of the GAT. Therefore, both the actor selecting the action and the critic computing the advantage of the action use the graph structure to learn the spatial features of the system's evolution due to both the demand pattern and the operator's matching decisions.

### B. AV-Request Assignment Approach

The approach for the AV-request assignment problem is adopted from [6] based on two applicable approaches:
  i. S1 sequentially assigns travelers FCFS to the nearest idle AV or relocating vehicle.
 ii. S2 solves the optimization problems presented previously to simultaneously determine the assignment of the requests to vehicles.

## IV. NUMERICAL EXPERIMENTS

To implement and evaluate the proposed approach, we use New York City (NYC) yellow taxi trip data made available by the NYC Taxi and Limousine Commission [37]. Specifically, we use the data from April 2016 for the island of Manhattan, which contain exact trip origin and destination coordinates that are not available in more recent data. Each taxi trip record additionally includes the passenger count, pick-up and drop-off date and time, and trip distance. Summary statistics from the data set are shown in Table II, computed after removing all zero-distance trips in the pre-processing stage.

TABLE II
NYC TAXI DATA SUMMARY STATISTICS FOR APRIL 2016

| | | Daily Trips | Hourly Trips | Trip Distance [km] |
|---|---|---|---|---|
| All Days | Mean | 329,838 | 13,743 | 2.84 |
| | St. dev. | 27,227 | 1,134 | 0.096 |
| | Min | 273,754 | 11,406 | 2.67 |
| | Max | 385,946 | 16,081 | 3.07 |
| Weekdays | Mean | 326,691 | 13,612 | 2.79 |
| | St. dev. | 22,653 | 943.8 | 2.67 |
| | Min | 273,754 | 11,406 | 2.84 |
| | Max | 367,412 | 15,308 | 0.04 |

The temporal distribution of the trips throughout the day is important to understanding the potential needs for fleet management. Figure 5 shows the number of hourly trips across each day in April 2016 and demonstrates distinct temporal peaks in demand during the day and differences between weekdays and weekends.

For these numerical experiments, we specify the service area as a rectangle around the island of Manhattan and divide the island into 16 approximately square sub-areas. The centroids for the sub-areas were computed as demand centroids, instead of area-based centroids. The simulation generates trip requests based on the observed taxi trip start times and corresponding pick-up and drop-off locations with a given demand fraction parameter. The demand fraction determines the rate at which demand is sampled from the data to generate simulation scenarios. This allows us to generate a large number of simulation episodes from the relatively small data set.

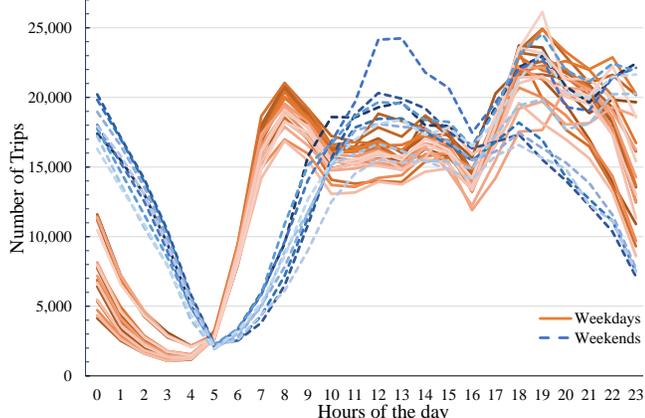

Fig. 5. With and between day variation in demand. Orange solid lines represent the weekdays, blue dashed lines represent the weekends.

### A. Experimental Design

The numerical experiments were designed to answer three research questions: (1) Does the ISR approach implicitly learn demand patterns and thus make demand forecasting superfluous? (2) Does ISR outperform repositioning-based repositioning approaches from the literature? (3) Does ISR transfer to scenarios not seen in training occurring due to changes in the agent's environment?

To answer the first question, we implement an alternative RLR approach that receives an input of future demand predictions as part of the system state, similar to [31]. This alternative approach will be referred to as externally guided repositioning (EGR). Its input demand predictions are based on the demand forecasting method in [19], modeling the rate of new user requests $\lambda$ in a Poisson demand generating process. Hence, we will compare the performance of ISR to that of EGR to evaluate our model's ability to bypass the need for demand predictions.

For the second question, the RLR approaches are compared to the joint optimization (JO) approach for matching and repositioning from [19], which uses the same demand prediction method mentioned above. The JO approach weighs the relative importance of repositioning and matching using assignment reward and imbalance penalty parameters. Due to the use of the demand fraction 0.1, we implement JO-1 with parameter values according to the original paper and JO-2 with parameter values scaled down with the fraction 0.1.

For the third question, we develop several scenarios with varying demand patterns and different assignment strategies, as characteristics of the operating environment for the RL agent. Table III shows the simulation parameter values and model settings. A total of 8 experiment scenarios are generated according to combinations of the following: two assignment strategies: FCFS (S1) and optimal (S2) assignment, two time periods: extended AM peak period (5 am to 1 pm) and full day (3 am to midnight), and two daily demand patterns for weekdays and weekends, as shown in Table IV. We select scenario 1 to be used for training and then test the RL agents on all eight scenarios to assess their performance on previously unseen scenarios.

TABLE III
SIMULATION PARAMETERS, VALUES, AND MODEL SETTINGS

| Parameter | Value | Applicable for Models: |
|---|---|---|
| Vehicle speed | 5 m/s (11 mph) | All |
| Drop-off time | 15 s | All |
| Pick-up time | 45 s | All |
| Simulation time step | 15 s | All |
| Demand fraction | 0.1 | All |
| AV fleet size | 600 | All |
| Inter-decision interval | 30 s | JO |
| Assignment interval | 30 s | ISR, EGR |
| Repositioning interval | 300 s | ISR, EGR |
| Prediction horizon | 30 min | JO-1, JO-2 |
| Prediction horizon | 90 min | EGR |
| Prediction interval | 300 s | EGR, JO-1, JO-2 |

TABLE IV
SPECIFICATIONS FOR 8 EXPERIMENTAL SCENARIOS

| Scenario | Period | Days | Assignment |
|---|---|---|---|
| 1 | AM Peak | Weekday | S1 (FCFS) |
| 2 | AM Peak | Weekday | S2 (optimal) |
| 3 | AM Peak | Weekend | S1 (FCFS) |
| 4 | AM Peak | Weekend | S2 (optimal) |
| 5 | Full Day | Weekday | S1 (FCFS) |
| 6 | Full Day | Weekday | S2 (optimal) |
| 7 | Full Day | Weekend | S1 (FCFS) |
| 8 | Full Day | Weekend | S2 (optimal) |

### B. Results and Analysis

The experimental results include tests of the ISR and its alternative EGR, along with two versions of the JO approach across the 8 testing scenarios. The main experiment results are presented in Table V. The system's performance is evaluated for service quality and fleet efficiency. Service quality is measured with the mean and standard deviation of passenger wait times. Fleet efficiency is evaluated via the empty distance travelled as a percentage of the total fleet distance travelled.

#### 1) Evaluation of Service Quality

In terms of service quality, the major observation is that RLR approaches lead to significant reductions in mean wait times, near 50% reduction relative to JO-1 in all but one scenario, and 60% reduction relative to JO-2 in all scenarios. These values highlight the superiority of RLR over JO in efficiently repositioning fleet vehicles to improve service quality. Between the two RLR approaches (namely, ISR and EGR), there are noticeable differences only for full day cases, i.e., scenarios 5 through 8, where EGR achieves additional wait time reductions ranging from 2.8% to 5.8% relative to ISR.

TABLE V
AVERAGE VALUES OF PERFORMANCE MEASURES

| | S | ISR | EGR | JO-1 | JO-2 |
|---|---|---|---|---|---|
| Mean request wait time (seconds) | 1 | 101 | 101 | 209 | 259 |
| | 2 | 100 | 99 | 208 | 259 |
| | 3 | 84 | 85 | 238 | 270 |
| | 4 | 84 | 85 | 238 | 273 |
| | 5 | 120 | 113 | 203 | 285 |
| | 6 | 109 | 103 | 202 | 285 |
| | 7 | 88 | 85 | 200 | 267 |
| | 8 | 87 | 84 | 200 | 273 |
| St. Dev. of request wait time (seconds) | 1 | 73 | 73 | 149 | 215 |
| | 2 | 73 | 70 | 149 | 215 |
| | 3 | 47 | 48 | 180 | 219 |
| | 4 | 47 | 48 | 181 | 220 |
| | 5 | 150 | 132 | 145 | 203 |
| | 6 | 112 | 96 | 145 | 203 |
| | 7 | 57 | 51 | 141 | 206 |
| | 8 | 55 | 50 | 142 | 210 |
| Percent empty distance | 1 | 53% | 53% | 49% | 25% |
| | 2 | 53% | 53% | 50% | 25% |
| | 3 | 66% | 66% | 44% | 25% |
| | 4 | 66% | 66% | 44% | 25% |
| | 5 | 53% | 52% | 44% | 24% |
| | 6 | 52% | 52% | 44% | 24% |
| | 7 | 54% | 54% | 43% | 23% |
| | 8 | 54% | 54% | 43% | 22% |

In terms of the standard deviation of wait times, RLR also yields significantly lower values compared to JO, more than 51% and 66% lower compared to JO-1 and JO-2, respectively, in all but two scenarios. In scenarios 5 and 6, however, RLR approaches exhibit higher standard deviation values relative other scenarios, which can be indicative of transferability issues. Specifically, these full-day cases with weekday demand patterns are the only scenarios exhibiting two major demand peaks, as seen in Figure 5. Therefore, we hypothesize that the presence of the second demand peak, not experienced in the model's training, can be the source of transferability issues. To evaluate this hypothesis and further analyze the variability of wait times, we compare the spatial distribution of wait times for the AM and PM peak periods. Specifically, we visualize the wait times across the service region for scenarios 5 and 7 in 6 and 7, respectively. Scenario 7 uses a full-day weekend demand pattern which, though unseen in training, does not include a second demand peak but rather has a delayed morning peak period that extends into the afternoon. For conciseness, we exclude the visualization for JO-2 where standard deviation values are consistently higher than those of JO-1. The graduated colors in Figures 6 and 7 represent the quintiles of observed wait times across the three models, determined separately for each scenario's AM and PM peak period.

Figure 6 shows that both RLR approaches exhibit significantly higher variation of wait times during the PM peak period (Fig. 6b and 6d) relative to the AM peak (Fig. 6a and 6c). On the other hand, the results in Figure 7 show more consistent performance for the AM and PM peak periods. This supports the idea that the increased wait time standard deviation in scenario 5 can be attributed to reduced performance in the PM peak period, which can be attributed to issues in transferability. While additional training on weekday PM peak period demands may be beneficial for this model, these results reveal the need for further understanding of the training needs for RL-based approaches in dynamic environments.

Some additional observations can be made from Figures 6 and 7 regarding the spatial distribution of variations in wait times. Specifically, we observe that lower wait times under the JO-1 approach exhibit distinct spatial clustering patterns, as seen in subfigures e) and f) in both Figures 6 and 7. This results in an imbalance of wait times between distinct neighborhoods, rather than an even distribution of high and low wait times as seen with the RLR approaches. While equity in the distribution of service quality may need to be further quantified and measured, these results indicate the RLR's potential for improving equity and fairness in SAMS. This is a particularly surprising finding, given that equitable distribution of vehicles or wait times was not part of the model's reward function. This observation requires further examination to understand if the results would hold across other metrics, and whether they would generalize or transfer to non-Manhattan networks.

### 2) Evaluation of Fleet Efficiency

Focusing on the fleet performance and efficiency, the results in Table V show an expected trade-off: improving passenger wait times comes at the cost of higher empty distance travelled by the fleet. Firstly, JO-2 exhibits the lowest percent empty fleet distance travelled, ranging from 22% to 25% across the 8 scenarios, while having the highest values for the mean and standard deviation of request wait times. JO-1 has the second-best values across all scenarios, ranging from 43% to 50% empty distance travelled, and was shown to have the second worst performance in terms of service quality metrics. In all but two scenarios, RLR approaches lead to a slight increase in empty fleet distance traveled compared to JO-1, ranging from 4 to 11 percentage points. However, in scenarios 3 and 4, their percent empty distance travelled with are higher by nearly 22 percentage points, a near 50% increase, relative to JO-1. This further illustrates the trade-off with service quality since scenarios 3 and 4 are ones where RLR approaches had the most significant wait time savings (65%) relative to JO-1.

For better understanding of the nature of empty vehicle travel, Figure 8 shows the breakdown of the percent empty distance travelled due to repositioning and passenger pick up. First, we observe that JO-2 exhibits exceptionally low repositioning distances, accounting for just 1-2% of the fleet's distance travelled. On the other hand, JO-1 performs

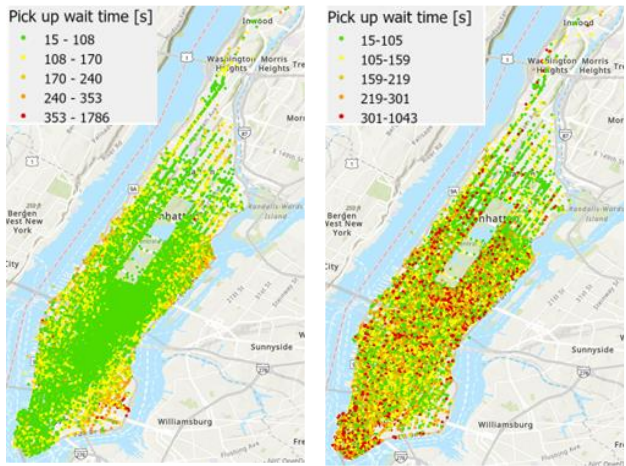

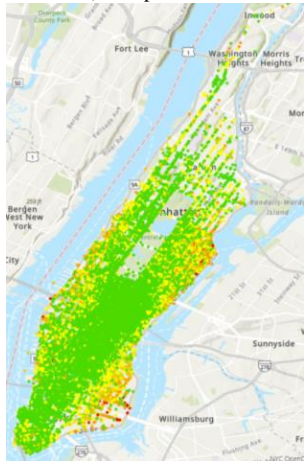
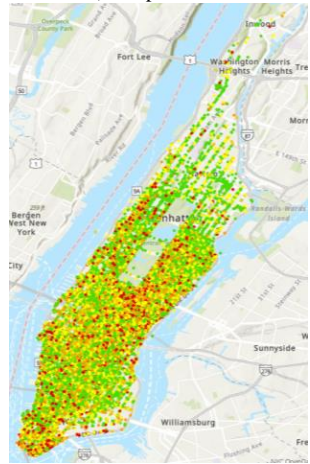

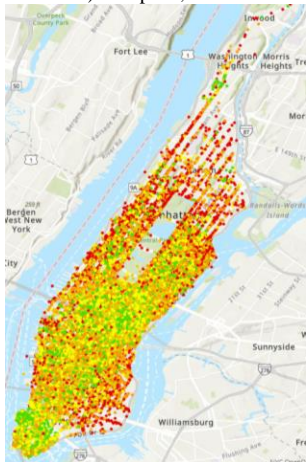
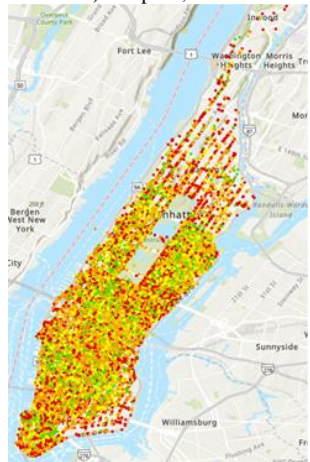

a) AM peak, ISR  b) PM peak, ISR

c) AM peak, EGR  d) PM peak, EGR

e) AM peak, JO-1  f) PM peak, JO-1

Fig. 6. Request wait times at pick up locations for scenario 5.

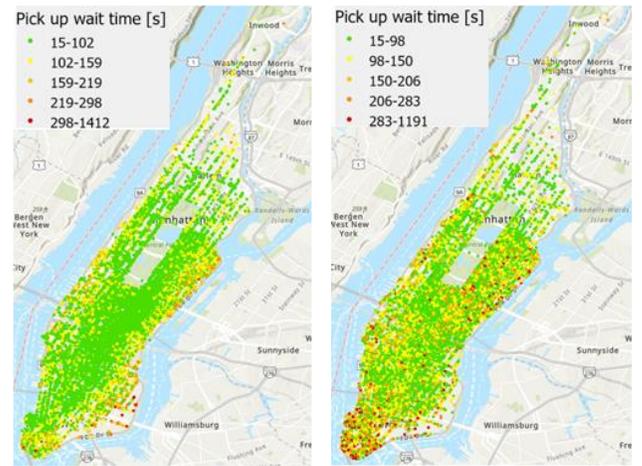

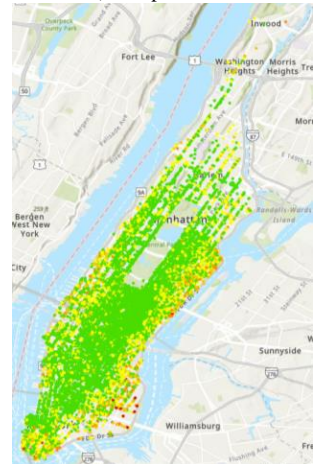
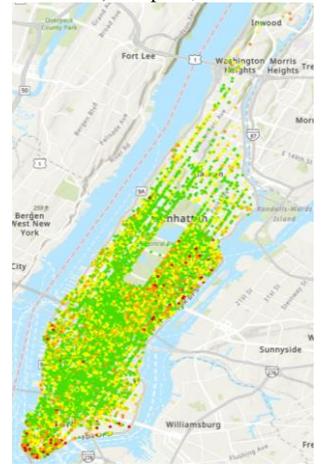

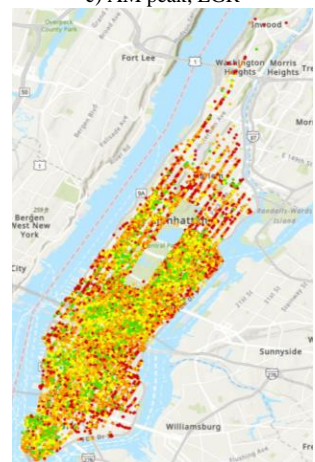
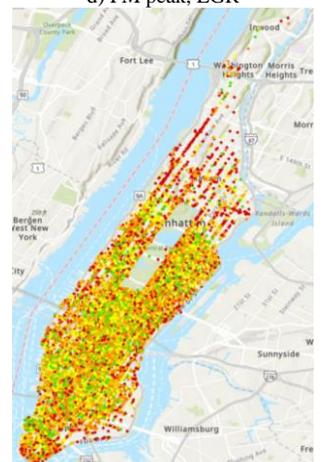

a) AM peak, ISR  b) PM peak, ISR

c) AM peak, EGR  d) PM peak, EGR

e) AM peak, JO-1  f) PM peak, JO-1

Fig. 7. Request wait times at pick up locations for scenario 7.

significantly more fleet repositioning, but does not reduce the percent empty pick-up distance relative to JO-2. With the RLR approaches, however, most of the empty distance is travelled in repositioning, resulting in significantly reduced empty pick-up distance. While contrasting the JO results and most other operational policies in the literature, where a higher proportion of empty miles stems from empty pick-up trips rather than repositioning, this result demonstrates the effectiveness of the RLR approaches. Namely, the remarkably short distances travelled during passenger pick up indicate that the vehicles are accurately placed close to upcoming demand. This result further supports the interpretation that the RLR's exceptionally low passenger wait times achieved are due to the vehicles' proximity to upcoming demand enabled by the repositioning strategy.

### 3) Overall Performance and Transferability

Across all three-performance metrics, we observe only negligible differences between the ISR and EGR models. The most significant differences are seen in full-day weekday cases (i.e., scenarios 5 and 6), for which we discussed the potential issues in transferability. This observation supports the interpretation that the model underperforms in presence of a PM peak period that was not observed in training. In those cases, the EGR has the additional advantage of receiving accurate externally generated demand forecasts to alleviate some of the transferability issues of the ISR. Further understanding is needed around choosing appropriate training scenarios to improve model transferability.

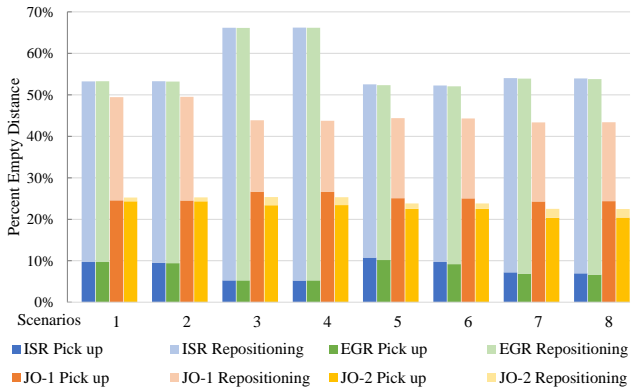

Fig. 8. Percent empty distance travelled due to passenger pick up (dark colors) and vehicle repositioning across testing episodes.

The experimental results indicate that the RLR approaches can significantly improve system performance compared to the JO approach. The gap in performance is particularly large in terms of service quality, where the learning-based model can achieve significant reductions in wait times and their standard deviation in most cases. In the context of real-world SAMS applications, especially in a competitive market, successfully reducing the pick-up distances and corresponding passenger waiting times can lead to increased market shares. Paired with an appropriate pricing strategy, the service provider can achieve a balance between the operational costs of additional empty miles and the users' willingness to pay for improved service quality. Nevertheless, the results raise a concern regarding the social cost repositioning if it generates significant amounts of excess travel in transportation networks. Thus, there are opportunities for future work to evaluate this trade-off and propose appropriate solutions to balance the cost of excess travel with the benefit of improved service quality.

## V. CONCLUSION

Focusing on the problem of fleet repositioning to improve the service quality and efficiency of SAMS fleet, this paper presents an RL-based repositioning (RLR) approach for integrated system-agent repositioning (ISR). This approach learns a fleet repositioning strategy in an integrated manner, bypassing the need for explicit demand predictions and cooperating with the operator's passenger-to-AV assignment decisions. The numerical experiments evaluate the ISR, comparing it to an externally guided repositioning (EGR) approach and a benchmark joint optimization (JO) method for passenger-to-AV assignment and AV repositioning.

The experiments demonstrate that the two RLR approaches significantly improve mean passenger wait times relative to the JO approaches, at the expense of increased percent empty fleet miles. Importantly, the effectiveness of the repositioning solution is evident in the significant decrease in the distance travelled for passenger pick-up resulting from appropriate positioning of AVs at locations and times conducive to quick passenger pick up. The ISR and EGR approaches exhibit comparable performance across most unseen scenarios. The results demonstrate the model's transferability to unseen demand patterns, extended operational periods, and changes in the AV assignment strategy. However, some challenges in transferability were uncovered, calling for further understanding of the types of system changes that may require retraining of RL models for SAMS decision-making.

This study opens opportunities for future research to extend the presented modeling approach. For example, it might be interesting to evaluate its performance with different spatial and temporal aggregation levels or pair the repositioning approach with different implicit or explicit demand forecasting models. The question of transferability presents opportunities for further research to evaluate the model's adaptability to unfamiliar scenarios and develop better understanding around the appropriate choice of efficient training sets. Future research can also consider problem variants incorporating different types of operator decisions, including shared rides, passenger hopping, and vehicle refueling or charging. Considering the observations related to equity in the distribution of service quality, this study raises some new questions in the domain of SAMS fleet operations. Specifically, it demonstrates potential for future work to augment the presented ISR approach for fair and equitable SAMS operation.